\documentclass[preprint,12pt]{elsarticle}
\usepackage{amssymb,amsmath}
\usepackage{tipa}
\usepackage{xcolor}

\journal{Computer Speech and Language}

\begin{document}

\begin{frontmatter}

\title{A study of vowel nasalization using instantaneous spectra}

\author{RaviShankar Prasad\dag \footnote{email: ravi.prasad@idiap.ch} \& B. Yegnanarayana\ddag}

\address{\dag Idiap Research Institute, Martigny, CH-1920 \\ \ddag International Institute of Information Technology, Hyderabad, India - 500032}
%\\ Running title : Co--articulatory nasalization : a production system perspective}

\begin{abstract}
Nasalization of vowels is a phenomenon where oral and nasal tracts participate simultaneously for the production of speech. 
Acoustic coupling of oral and nasal tracts results in a complex production system, which is subjected to a continuous changes owing to glottal activity. 
Identification of the duration of nasalization in vowels, and the extent of coupling of oral and nasal tracts, is a challenging task. 
The present research focuses on the changes in instantaneous vocal tract system response to study the effects of co--articulatory load of nasals on vowels. 
The dominant resonance frequency (DRF) contour derived from the spectra illustrates the contribution of oral and nasal resonances during nasalization. 
The extent of coupling determines the dominance of these resonances during glottal open and closed phases. 
A higher extent leads to more decay of energy for the oral resonances, and hence the vowel spectra exhibits a dominant nasal resonance. 
A comparison of the proposed parameters is made with the previously suggested measures. 
Several examples of segments of vowels in the context of nasal consonants for English language for both male and female speakers of English are used to illustrate different aspects of the proposed analysis method. 
\end{abstract}

\begin{keyword}
nasalized vowels; oral--nasal coupling; zero time windowing; HNGD spectra; dominant resonance frequency

\end{keyword}

\end{frontmatter}

%% \linenumbers

%% main text
\section{Introduction}
\label{section:sec1}
Nasalization of vowels occurs when the air flow takes place through both oral and nasal tracts during production of vowels. 
Oral vowels are produced by exciting an open oral tract with pseudo--periodic vibration of the vocal folds at the glottis. 
Nasal consonants are produced with lowering the velum and closing the oral tract, resulting in airflow only through the nasal tract. 
Coupling of an open oral and nasal tracts results in presence of a nasal signature in the vowel spectrum. 
Extent of this coupling dictates the volume of airflow through each of these tracts. 
The nasal tract consists of multiple sinus cavities, and acts as a branched cavity to the open oral tract during nasalization. 
This overall system is further subjected to glottal activity, resulting in a complex production system changing continuously with time. 
A dynamic and involuntary nature of glottal and velar movement makes the identification of presence of nasalization in vowels a challenging task. 
Understanding the phenomena of vowel nasalization in speech is important for the improvement of several speech applications, such as automatic speech recognition, speech pathology, language identification, and interactive voice response systems. 

Study of the phenomenon of vowel nasalization has been motivated to highlight the contrast between oral and nasal vowels. 
Previous studies explored the spectro--temporal characteristics of vowel segments derived across a variety of acoustic and linguistic contexts. 
Early developments towards the task have highlighted the presence of a low frequency spectral peak (within $250$--$300$ Hz range) in vowel spectra, as a characteristic signature of coupling of nasal and oral tracts \cite{house1956analog, fant1958acoustic, hawkins1985acoustic, glass1985detection}. 
These studies also emphasized on the role of spectral zero in $700$--$1800$ Hz range, and widening of the first formant ($F$\emph{1}) bandwidth, as supportive evidences of nasalization in vowels. 
Presence of a pole--zero pair around first formant ($F$\emph{1}) of the vowel, and a spectral peak around $1$ kHz, is also noted for several cases of nasalized vowel segments \cite{huffman1990implementation}. 
The effect of introduction of a synthetic pole--zero pair in low frequency range, and increasing the amplitude of the first harmonic, is studied for its contribution towards perception of nasality in vowels \cite{house1956analog, klatt1990analysis}.
Further studies have discussed the importance of spectral peak in $250$--$450$ Hz range as an important cue to determine the presence of nasalization in vowel spectra \cite{fujimura1971sweep, lindqvist1976acoustic, baavegaard1993vocal, dang1994morphological}.

These studies have propelled the task of automatic identification of nasalization in speech, based on parameters derived from vowel spectra. 
Spectral correlates $A$\emph{1}--$P$\emph{1} and $A$\emph{1}--$P$\emph{0} have been proposed to highlight the contrast within oral and nasalized vowels  \cite{chen1995acoustic, chen1997acoustic}. 
$A$\emph{1} is defined as amplitude of the harmonic peak closest to an estimated location of $F$\emph{1}. 
$P$\emph{1} is amplitude of the nasal peak in vicinity of $F$\emph{1}, whereas $P$\emph{0} is amplitude of first resonance peak at low frequencies. 
A large shape of the nasal tract contributes towards broadening of $F$\emph{1}, consequently lowering its amplitude in nasalized vowels. 
An average value of both these correlates were found to be lower for nasalized vowels than for oral vowel segments. 
The study presents a comparison between nasalized vowels for anticipatory (occurring after) and carryover (occurring before) nasal contexts, for these correlates. 
An \emph{`orality threshold'} measure is proposed based on the extent of lowering of velum, to distinguish nasalized vowels from oral vowels \cite{huffman1990implementation}. 
However, a high degree of variability in the nasalization phenomena across utterances, speakers and languages mitigates weakens the reliability on such a measure \cite{ploch1999nasals}. 
A set of nine acoustic features have been proposed to detect the presence of nasalization in vowel segments \cite{pruthi2007acoustic}. 
These features capture the low frequency spectral behavior of nasalized vowels, such as change in energy and bandwidth of $F$\emph{1}, and other spectral peaks, and changes in energy profile of different bands in low frequency range. 
All these parameters are derived on the short--time spectral representation of vowel spectrum obtained using short time Fourier transform (STFT). 
Dominant peaks in group delay (GD) spectrum have been utilized to detect hypernasality in pathological speech \cite{vijayalakshmi2007acoustic}. 
The GD spectrum is derived as derivative of the phase component of STFT with respect to frequency. 
Advantages of the GD spectrum is that it provides sharper spectral peaks compared to the STFT based spectrum \cite{bayya78jasa}.

Importance of the low frequency resonance occurring below or in the range of $F$\emph{1}, for identification of nasalization in vowels, has consistently been highlighted by several studies. 
The present study utilizes this knowledge to derive correlates from the instantaneous spectra to address the identification of presence of nasalization, along with its duration and extent, in vowel segments. 
The spectral representation obtained from the zero time windowing (ZTW) method gives a good resolution in temporal and spectral domains \cite{bayya2013spectro}. 
The method, therefore, proves helpful to study the dynamic characteristics of the production system during coupling of the oral and nasal tracts, subjected to the glottal activity. 
The paper explores the effects of contextual load of nasals on vowels, for different CV/VC pairs in the English language. 
The paper is organized as follows: 
Section \ref{backgroundANDmotivation} presents the zero time windowing method, and the motivation to the study.  
Section \ref{drfAnalysisoralANDnasalized} discusses the analysis of oral and nasalized vowels using dominant resonance frequencies (DRFs) derived from the instantaneous spectra. 
Section \ref{variabilityOFnasalCoupling} illustrates the behavior of DRF contour for different extent of coupling of oral and nasal tracts. 
Section \ref{observationCVCboundaries} discusses the observations based on proposed hypothesis for the behavior of DRF contour for oral and nasalized vowel segments for several CV/VC pairs. 
Section \ref{summary} presents a summary to the paper. 
%%%%%%%%%%%%%%%%%%%%%%%%%%%%%%%%%%%%%%%%%%%%%%%%%%%%%%%%%%%%%%%%%%%%%%%%%%%%%%%%%%%%%%%%%%%%%%%%%%%%%%%%%%%%%%%%%%%%%%%%%%%%%%%%%%%%%%%%%%%%%%%%%%%%%%%%%%%%%%%%
\section{Background and motivation}
\label{backgroundANDmotivation}
This section discusses the zero time windowing method to obtain the spectral characteristics at good temporal resolution.
Distinction in the dominant spectral behavior of oral vowels and nasal consonants provide the necessary motivation to explore vowel nasalization phenomena using ZTW.
\subsection{Zero time windowing method}
\label{section:zeroTimeWindowing}

The zero time windowing (ZTW) method utilizes a heavily decaying window, giving more weightage to samples near the point of application of the window, called the zero time \cite{bayya2013spectro}. 
This windowing is motivated by the zero frequency filter (ZFF) which is a sharp resonator centered around $0$ Hz \cite{murty2008epoch}. 
The time domain analog of this operation is a heavily decaying window function given by,
\begin{equation}
w_1[n] = \begin{cases} 0, \quad \emph{n} = 0, \\ 1/4sin^2(\pi n/2N), \quad \emph{n} = 1, 2, \ldots, \emph{N}-1, \end{cases} 
\label{eqn1}  
\end{equation}
where $N$ is the length of the window in samples corresponding to a duration of $l$ ms. 
The windowed signal is given as $x[n] = s[n]w[n]$, where $s[n]$ is the speech signal, and the window function is given as $w[n] = w_1^2[n]w_2[n]$. 
$w_2[n]$ is another window which helps to reduce the ripple effect due to truncation, and is given by 
\begin{equation}
w_2[n] = 4cos^2(\pi n/2N), \quad n = 0, 1, \ldots, N-1,  
\label{eqn2}
\end{equation}

Application of the window function $w_1[n]$ can be interpreted as an integration operation performed twice in the frequency domain \cite{bayya2013spectro}. 
Spectral characteristics of the windowed segment $x[n]$ are thus obtained by successive differentiation. 
The spectrum is represented using the Hilbert envelope of the differenced numerator group delay (HNGD) function. 
The numerator of GD (NGD) function is given by, 
\begin{equation}
 \tau(\omega) =  X_R(\omega)Y_R(\omega)+X_I(\omega)Y_I(\omega). %{X_R(\omega)^2+X_I(\omega)^2}
 \label{ngd_eqn}
\end{equation}
where $X(\omega)$=$X_R(\omega)$+$jX_I(\omega)$ is the discrete--time Fourier transform (DTFT) of $x[n]$, and $Y(\omega)$=$Y_R(\omega)$+$jY_I(\omega)$ is the DTFT of $nx[n]$. 
The Hilbert envelope of the twice differenced NGD (HNGD) shows the formant peaks with a good resolution \cite{anand2006extracting}. 
A window with duration $l\leq$ average pitch period gives the changes in the acoustic system response within a glottal cycle \cite{rspJASA}. 
The window $w[n]$ is shifted by one sample, to obtain the spectrum at every sampling instant. 

\begin{figure}
\centering
\includegraphics[height = 7cm, width=\linewidth, trim=1cm 1cm 0cm 5cm]{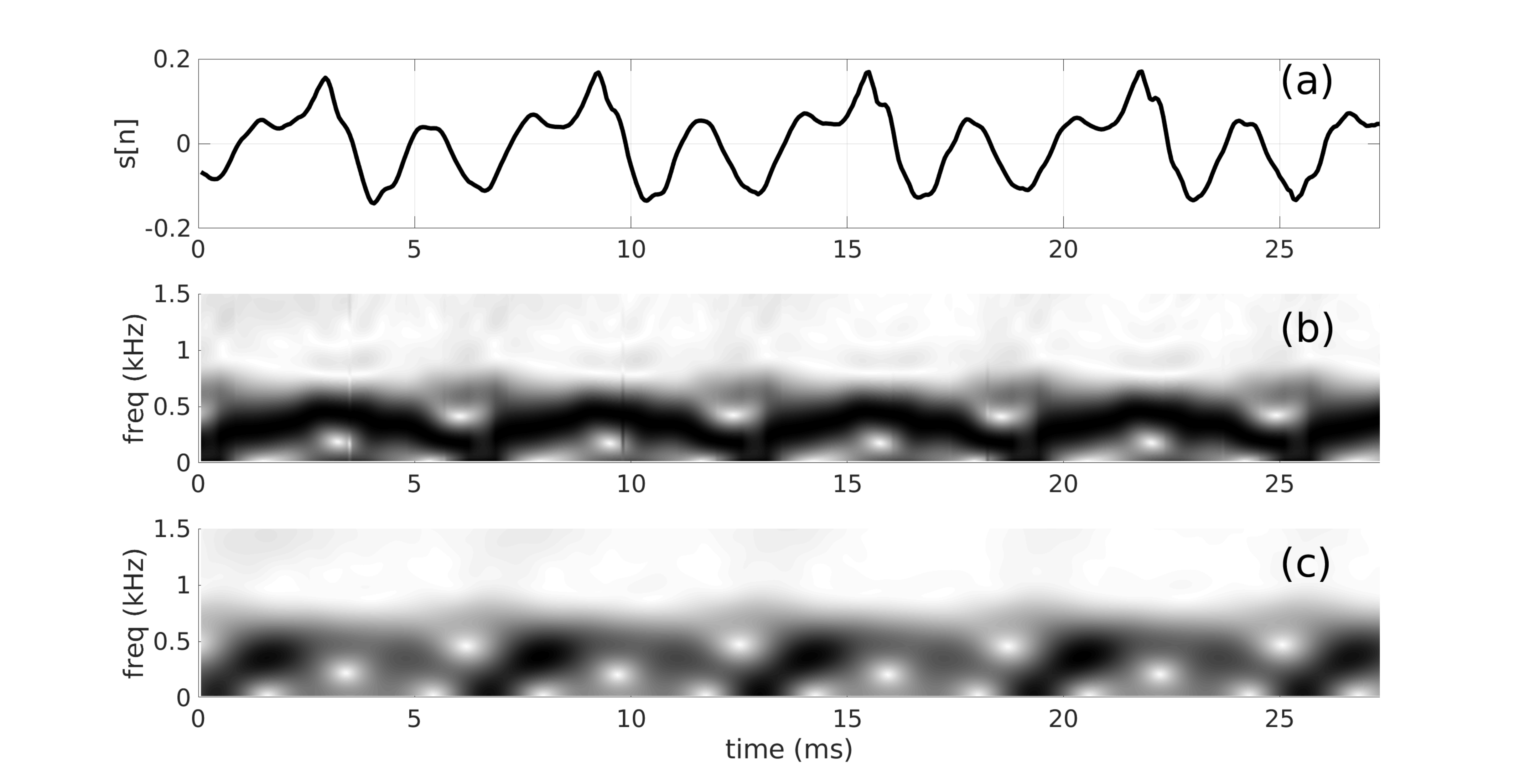} % Figure2
\caption{{Estimated spectrograms for a vowel segment using ZTW and STFT methods. 
(a) Segment of speech signal.
(b) HNGD spectrogram using ZTW analysis method with $l=4$ ms.  
(c) STFT spectrogram using a \emph{Hann} window of $4$ ms.}}
\label{figure:fig1}
\end{figure}

Figure \ref{figure:fig1} illustrates the contrast in spectral estimates obtained using HNGD and STFT methods.  
Figure \ref{figure:fig1}(b) shows the HNGD spectrogram for a segment of vowel /\textipa{e}/ uttered by a male speaker given in Fig. \ref{figure:fig1}(a). 
The HNGD spectrogram is obtained across an analysis segment duration $l=4$ ms, shifted at every sample. 
Changes in the system response appear in an instantaneous manner in the HNGD spectrum. 
Figure \ref{figure:fig1}(c) shows the STFT spectrogram of the segment. 
The STFT is computed at every sampling instant using $l=4$ ms \emph{Hann} window. 
The averaging effect on the spectral details can be seen in the STFT spectrogram. 
Movement of the spectral resonances can be seen better in the HNGD spectrogram. 

\subsection{Dynamics of coupling of oral and nasal tracts}
\label{section:sec3}
The phenomenon of nasalization is categorized into three major types \cite{pruthi2007analysis}: 
\begin{itemize}
 \item \textbf{Co--articulatory nasalization:} The cases where nasal consonants appear either in the pre--vocalic or post--vocalic or both contexts with a vowel. 
 Production of a nasal requires complete closure of the oral tract. 
 Delay in the closure of the oral tract, while opening the velopharyngeal section, leads to nasalization of the vowel preceding a nasal consonant. 
 On the other hand, opening of the oral tract before closure of the velopharyngeal section leads to nasalization of the vowel following a nasal consonant. 
 \item \textbf{Phonemic nasalization:} The cases where vowels are distinctively nasalized, independent of any contextual proximity to nasals. 
 Oral and nasalized vowels form a minimal pair for such cases, which are linguistically different from each other and therefore convey different meaning. 
 \item \textbf{Functional nasalization:} The cases where characteristics of nasalization are introduced in vowels due to dysfunctional velar mechanism. 
\end{itemize}

Co--articulatory nasalization is the most dynamic of the three types of nasalization discussed above, in terms of production system complexity, duration, dependence on factors related to contextual load, muscular inertia, articulatory constraints, speaking rate etc. 
Vowels appearing in context of nasal consonants are expected to be nasalized, at least for part of their duration.
This assumption may hold true across different utterances and speakers in different languages. 
Furthermore, the degree of nasalization may also vary for different utterances of same VC/CV pair. 

Recent studies show that acoustic correlates $A$\emph{1}--$P$\emph{1} and $A$\emph{1}--$P$\emph{0} are popular in detecting the presence of nasalization in vowels \cite{styler2017acoustical}. 
However, derivation of these parameters requires manual demarcation of locations of $F$\emph{0}, $P$\emph{0} and $P$\emph{1} in STFT, which is a tedious process. 
Furthermore, the harmonic structure of spectrum and the analysis window response pose difficulty in resolving the nasal formant and other spectral peaks in STFT. 
The dominant resonance frequency (DRF) contour derived from the HNGD spectra are therefore utilized to identify the presence of nasal resonance in vowel spectra. 
DRFs have proven to be a consistent and concise representation of the spectral characteristics in speech, and efficiently reflect changes in production system due to coupling/decoupling of cavities to the oral tract \cite{rspJASA}. 
Ability of DRFs to capture distinction in the production system response for oral and nasalized vowels also serves as motivation to this study.

%%%%%%%%%%%%%%%%%%%%%%%%%%%%%%%%%%%%%%%%%%%%%%%%%%%%
\section{Analysis of oral and nasalized vowels using DRFs}
\label{drfAnalysisoralANDnasalized}
The section describes behavior of DRF contours for oral vowels and nasal segments. 
The behavior is further investigated for nasalized vowel segments. 
Previous studies hypothesized that vowels present in these contexts are always nasalized. 
The present section investigates the behavior of DRFs in oral vowels and nasal segments. 
A hypothesis is suggested on the basis of this behavior, to study the behavior of DRF contour during vowel nasalization. 
This hypothesis is validated based on previously proposed spectral correlates. 
The validation is conducted for vowels in English language, obtained from utterances recorded by male and female speakers of English language in TIMIT database \cite{TIMIT}. 

\subsection{Distinction in behaviour of DRFs for oral vowel and nasal segments}
\label{drfBehavior_vowelNasal}
Distinction in the spectral structure among nasal and oral vowel segments has been widely discussed in literature.  
Presence of a characteristic resonance in lower frequencies, and a spectral null following it, is attributed to the presence of a longer nasal cavity coupled with a closed oral cavity, during production of nasals. 
Production of oral vowels is usually characterized by spectral resonances in relatively higher frequency range. 
These spectral characteristics are derived at a frame duration $20$--$30$ ms with an underlying assumption of stationarity of the production system. 
Such analysis averages the behavior of speech production system, which otherwise is continuously changing in nature. 
%%%%%%%%%%%%%%%%%%%%%%%%%%%%%%%%%%%%%%%%%%%%%%%%%%%
\begin{figure}[htb]
\centering
\includegraphics[height = 7.5cm, width=\linewidth]{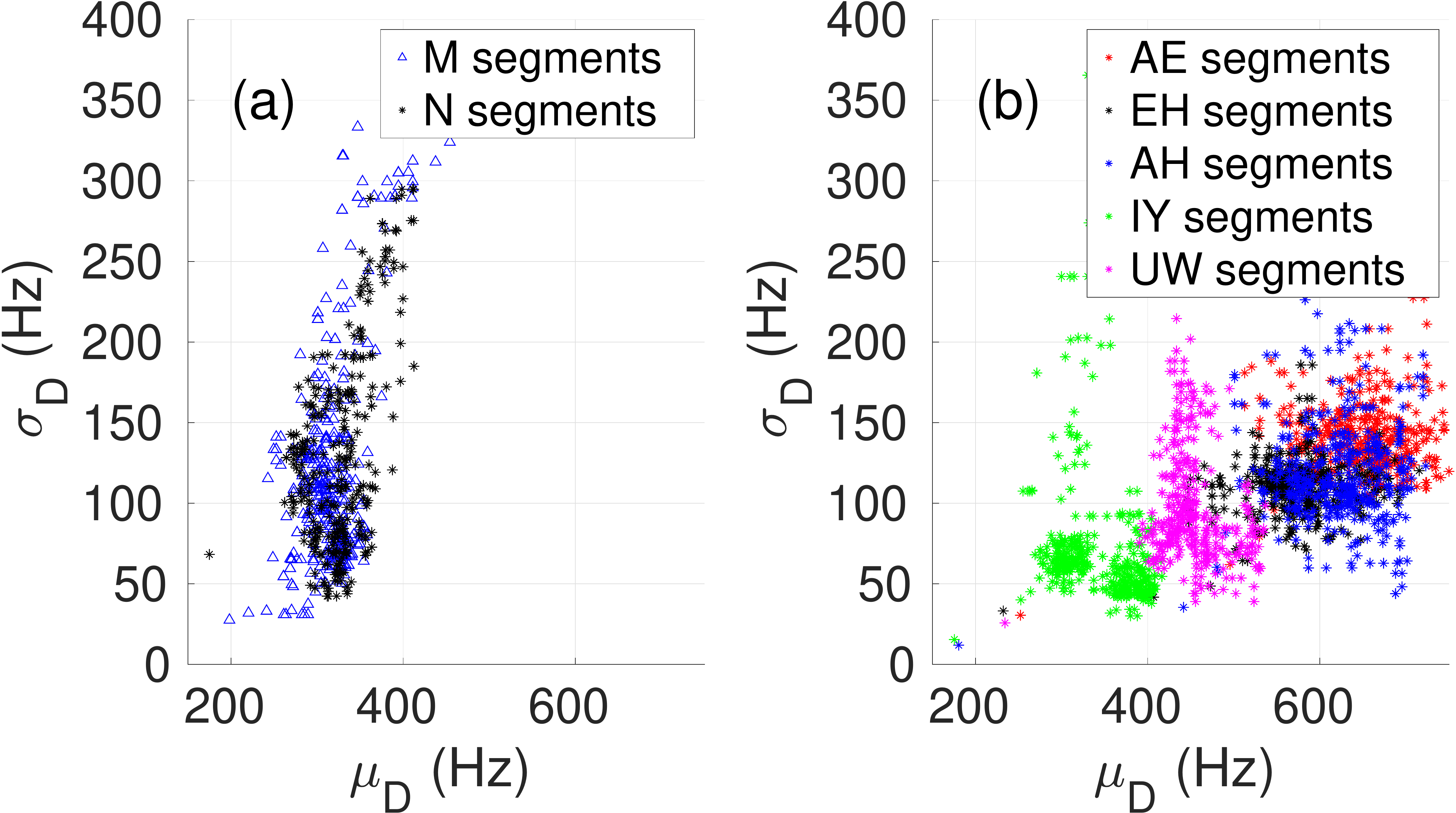}
\caption{$\mu_D$ vs $\sigma_D$ across glottal cycles for (a) nasal consonants /m/ and /n/, and (b) vowels /\textipa{\ae}/, /\textipa{e}/, /\textipa{2}/,   /i/, and /\textipa{u}/ . 
} 

\label{figure:fig3}
\end{figure}
%%%%%%%%%%%%%%%%%%%%%%%%%%%%%%%%%%%%%%%%%%%%%%%%%%%

The present study uses HNGD spectrum, obtained over short duration of $4$ ms, to derive the DRF contours reflecting instantaneous changes in production system response. 
The glottal activity during production of oral and nasal vowels leads to a fluctuation of DRFs within high and low frequency range for each glottal cycle \cite{rspJASA}. 
These fluctuations are characterized using mean ($\mu_D$) and standard deviation ($\sigma_D$) derived over the DRF contour across every glottal cycle, for vowel and nasal segments. 
$\mu_D$ indicates the  centroid and $\sigma_D$ indicates the bounds of fluctuation in DRF contours for respective utterances. 
Fig. \ref{figure:fig3} shows the distribution of these values in spectral plane, obtained over multiple instances of oral vowel and nasal segments in continuous speech in English language, for utterances chosen from TIMIT dataset \cite{TIMIT}. 
Fig. \ref{figure:fig3}(a) gives location of $\mu_D$ vs. $\sigma_D$ values for nasal (/m/ and /n/) segments, and Fig. \ref{figure:fig3}(b) gives these for oral vowel (/\textipa{\ae}/, /\textipa{e}/, /\textipa{2}/,   /i/, and /\textipa{u}/) segments. 
Oral vowels are chosen as vowel segments appearing in context with fricative and stop consonants. 
Except for /i/, the non--overlapping clusters for oral vowels and nasals show distinction in the production system characteristics for oral vowels and nasals. 
Literature have highlighted the difficulties in identifying nasalization within front vowels \cite{chen1995acoustic}. 
Thus, the study will not cover instances of /i/ for analysis.  
For other oral vowels and nasals, clusters of $\mu_D$ vs. $\sigma_D$ help in deriving bounds for the expected range of occurrence and fluctuations of DRF contours. 
Given the figure, the $300$--$400$ Hz range (= $B_N$) is understood as expected range of fluctuation of DRF contour for nasal segments (Fig. \ref{figure:fig3} (a)), and the $450$--$850$ Hz range (= $B_V$) for oral vowels (Fig. \ref{figure:fig3} (b)). 

%%%%%%%%%%%%%%%%%%%%%%%%%%%%%%%%%%%%%%%%%%%%%%%%%%
\begin{figure}[htb]
\centering
\includegraphics[height = 8cm, width=\linewidth]{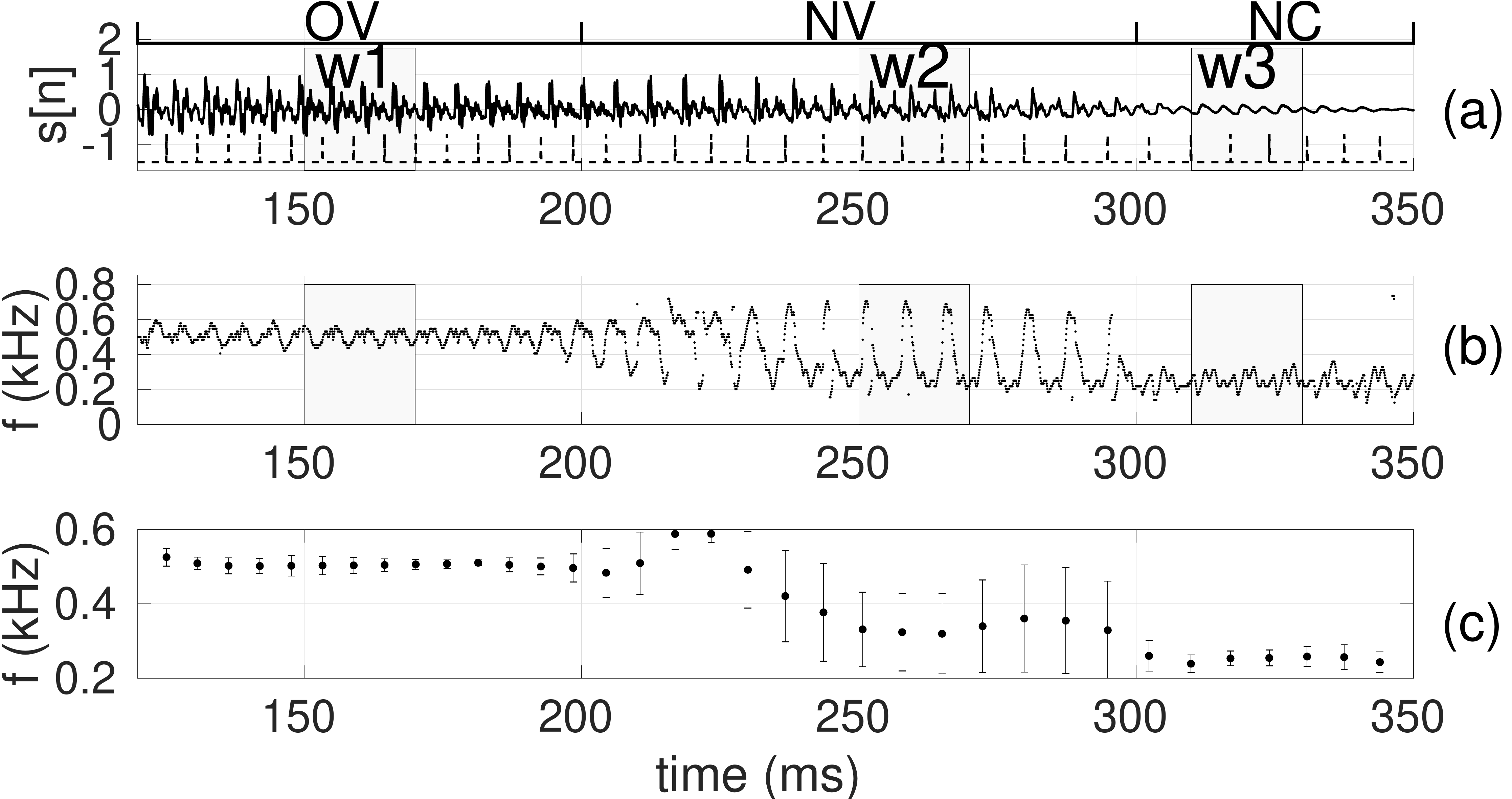}
\caption{DRF contour obtained for VC transition segment. 
(a) Speech with the GCI locations (dotted). 
(b) DRF contour.
(c)  $\mu_D$ (dots) and $\sigma_D$ (bars) within each glottal cycle.} 
\label{figure:figNasalizedVowelDRF}
\end{figure}
%%%%%%%%%%%%%%%%%%%%%%%%%%%%%%%%%%%%%%%%%%%%%%%%%%

%%%%%%%%%%%%%%%%%%%%%%%%%%%%%%%%%%%%%%%%%%%%%%%%%%%%%%%%%%%%%%%%
\begin{figure}[htb]
\centering
\includegraphics[height = 8cm, width=\linewidth]{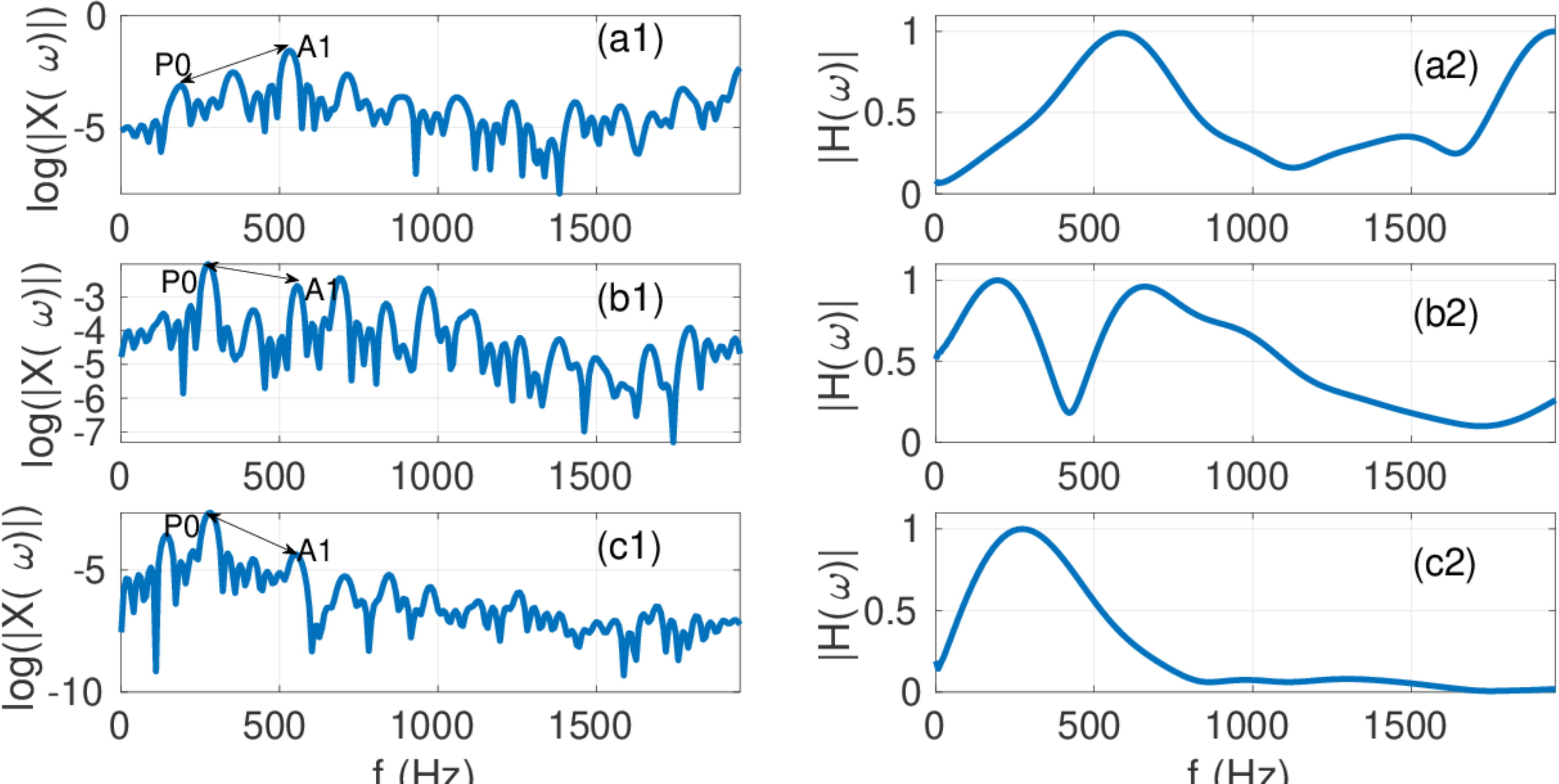}
\caption{Comparison of DFT and HNGD spectra for OV, NV and NC segments. 
(a1), (b1) and (c1) DFT spectra in OV, NV and NC regions, respectively, with the $A$\emph{1}--$P$\emph{0}.
(a2), (b2) and (c2) HNGD spectra. 
} 
\label{figure:figDFTandHNGD_oralNasal}
\end{figure}
%%%%%%%%%%%%%%%%%%%%%%%%%%%%%%%%%%%%%%%%%%%%%%%%%%%%%%%%%%%%%%%%
\subsection{DRF contour for vowels appearing in nasal context}
\label{drfs_vowel_nasalized}
Vowels appearing in CV/VC context with nasal consonants experience a contextual load, which dictates the graduation of instantaneous spectral behavior from  $B_V$ to  $B_N$, towards the conjunction boundary of these segments. 
A sudden transition reflects the absence of overlap of their respective spectral characteristics. 
A gradual transition of DRFs within $B_V$, towards the characteristics of $B_N$, within the vowel segment indicates an overlap of nasal spectral characteristics. 
Fig. \ref{figure:figNasalizedVowelDRF} illustrates one such example of graduation in DRF contour across the length of a vowel segment, present with a nasal context. 
Fig. \ref{figure:figNasalizedVowelDRF}(a) shows the signal for a VC pair /\textipa{\ae}/ and /n/, in the utterance \emph{`man'} spoken by a female speaker of English language.  
The DRF contour obtained from the HNGD spectra using the ZTW method with an analysis window of $4$ ms is shown in Fig. \ref{figure:figNasalizedVowelDRF}(b). 
The VC boundary appears at $300$ ms as given by the annotations provided with the database. 
Changes in the behavior and fluctuations of DRF contour across the length of vowel can be noted in the figure. 
Glottal cycles are identified using glottal closure instants (GCIs) derived using ZFF method \cite{murty2008epoch} (Fig. \ref{figure:figNasalizedVowelDRF}(a)).

Three locations, $w1$ (near the vowel onset), $w2$ (near the VC conjunction region), and $w3$ (post the conjunction region) are chosen to illustrate the distinction and similarity, in spectral behavior across the vowel segment with the nasal spectra, for different instances. 
The locations, $w1$ and $w2$ lie in the vowel segment whereas $w3$ lies in the nasal segment. 
Figs. \ref{figure:figDFTandHNGD_oralNasal}(a1), \ref{figure:figDFTandHNGD_oralNasal}(b1) and \ref{figure:figDFTandHNGD_oralNasal}(c1) show the DFT spectrum obtained at $w1, w2$ and $w3$ in the segment. 
The oral formant can be resolved around $550$ Hz as the strongest harmonic peak, in Fig. \ref{figure:figDFTandHNGD_oralNasal}(a1), with an amplitude value understood as $A$\emph{1}. 
The characteristic low frequency nasal resonance is resolved around $300$--$350$ Hz, in Fig. \ref{figure:figDFTandHNGD_oralNasal}(c1), with an amplitude value $P$\emph{0}. 
The parameter $A$\emph{1}--$P$\emph{0} exhibits relatively a higher value at the window location $w1$ (Fig. \ref{figure:figDFTandHNGD_oralNasal}(a1)) as compared to location $w2$ (Fig. \ref{figure:figDFTandHNGD_oralNasal}(b1)). 
The value of $P$\emph{0} appears larger than $A$\emph{1} for the case of location $w2$, which is closer to the VC conjunction region, resulting in a negative value of the parameter. 
Another limitation with the parameter $A$\emph{1}--$P$\emph{0}, apart from resolving the locations of spectral peaks, is that it needs to be observed in a relative frame of reference. 
It is therefore difficult to determine the presence of nasalization for segments with few instances of the same VC/CV pair. 

Figs. \ref{figure:figDFTandHNGD_oralNasal}(a2), \ref{figure:figDFTandHNGD_oralNasal}(b2) and \ref{figure:figDFTandHNGD_oralNasal}(c2), show the HNGD spectrum obtained at $w1, w2,$ and $w3$ using $l=5$ ms, respectively. 
The oral and nasal resonances in $550$--$600$ and $300$--$350$ Hz regions are easily resolved in the HNGD spectra in Figs. \ref{figure:figDFTandHNGD_oralNasal}(a2) and \ref{figure:figDFTandHNGD_oralNasal}(b2), respectively. 
The HNGD spectrum at $w2$ exhibits both oral and nasal resonances with similar dominant behavior. 
This asserts the simultaneous presence of both these cavities during the production, which is also verified with a relatively lower value of $A$\emph{1}--$P$\emph{0} in this region. 
The DRF contour in Fig. \ref{figure:figNasalizedVowelDRF}(b) illustrates the distinction in oral and nasalized segments based on the bound of their fluctuation within each glottal cycle. 
The oral vowel and nasal segments exhibit DRFs in $B_V$ and $B_N$ ranges, respectively, as suggested by the $\mu_D$ and $\sigma_D$ values in Fig. \ref{figure:figNasalizedVowelDRF}(c).  
The nasalized segments exhibit DRFs fluctuating between $B_V$ and $B_N$ ranges, owing to presence of a coupled oral and nasal tract, leading to a shift in $\mu_D$ and higher $\sigma_D$ values.  
Due to a high temporal resolution, it is easier to demarcate the presence of nasalization for each glottal cycle with an improved accuracy.

\section{Behavior of DRF contour for a variability in coupling of oral and nasal tracts}
\label{variabilityOFnasalCoupling}
%%%%%%%%%%%%%%%%%%%%%%%%%%%%%%%%%%%%%%%%%%%%%%%%%%%%%%%%%%%%%%%%%%%%%%%%%%%%%%%%%%%%%%%%%%%%%%
DRFs appear as distinctive feature of the HNGD spectrum to represent the instantaneous vocal tract system response. 
Adduction of the nasal tract to the oral tract results in the dominance of a characteristic low frequency nasal resonance, which is efficiently captured by the DRFs in HNGD spectrum. 
The presence section studies the characteristics of DRF contours for different cases of oral--nasal coupling.  
The effect of a higher extent of this coupling on DRF behavior is explored in detail. 
The section also highlights the importance of segment duration, and hence determination of the glottal open phase, towards identification of nasalization in vowels. 

\subsection{Behavior of DRF for vowels nasalized at different extents}
\label{section:sec4.1}
A significant transition in the behavior of DRFs in vowel segments is illustrated by deviation in their fluctuation from $B_V$ range. 
There can be several factors to cause such a deviation, but for the study of nasalization in vowels, such a behavior is largely dependent on the extent of coupling of the oral and nasal tracts. 
Presence of nasalization can easily be tracked using the $\mu_D$ and $\sigma_D$ parameters derived for the DRF contours. 
A change in the extent of coupling, however,  dictates the extent of shift in values of these parameters. 

\begin{figure}
\centering
\includegraphics[height = 7.8cm, width=\linewidth, trim=0cm 0cm 0cm 5cm]{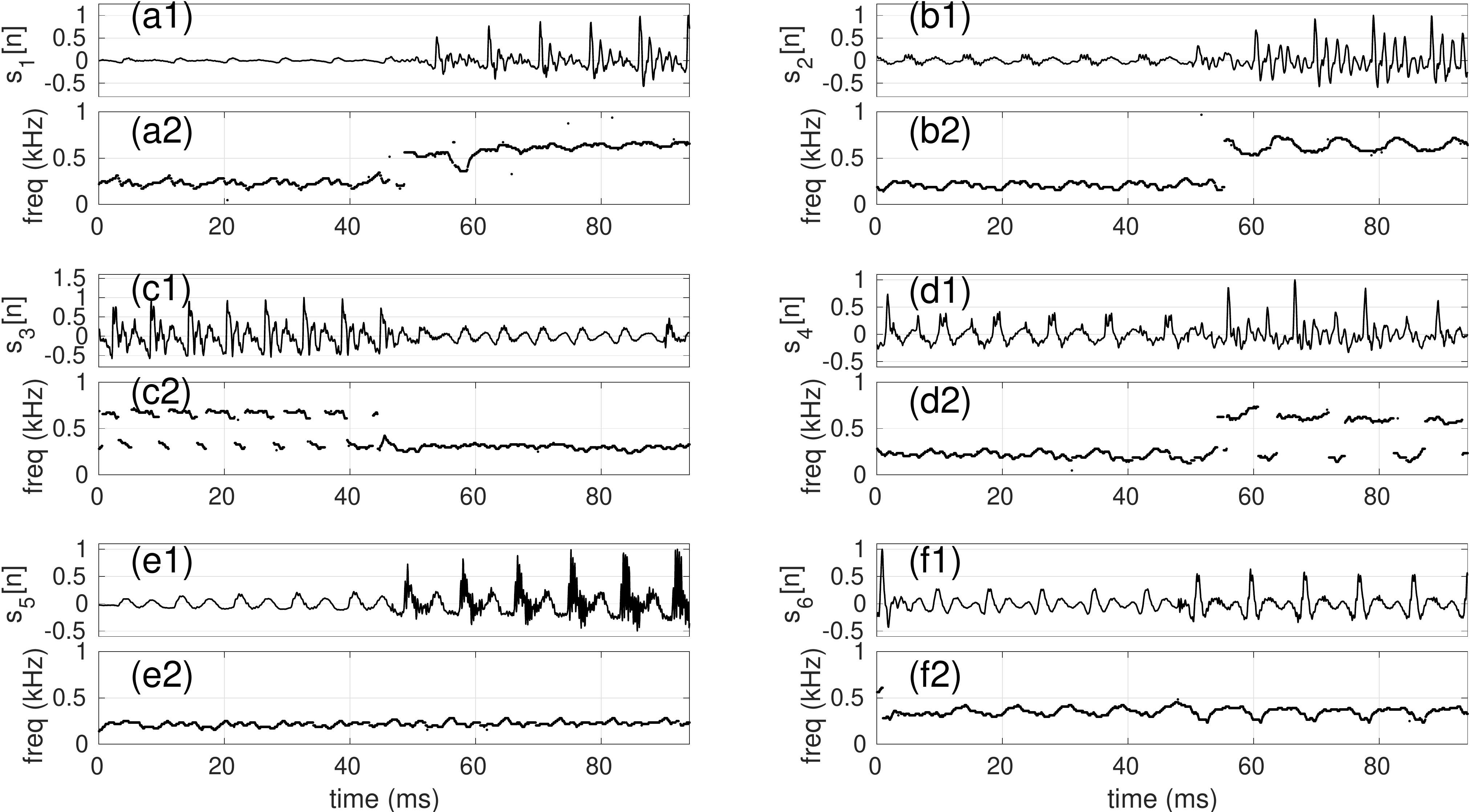} %Figure4   Report_Figure_5_New Report_Figure_5_differentExtent
\caption{DRF contours obtained using ZTW analysis, for different extents of nasalization. 
(a1)--(f1) Speech signals for syllable \emph{/}\textipa{me}\emph{/}, \emph{/}\textipa{m2}\emph{/}, \emph{/}\textipa{\ae n}\emph{/}, \emph{/}\textipa{m2}\emph{/}, \emph{/}\textipa{mi}\emph{/} and \emph{/}\textipa{n1}\emph{/}, respectively.
(a2)--(f2) DRF contours for the corresponding signals.}
\label{figure:DRFs_Nasal_No_Part_Full}
\end{figure}

Figure \ref{figure:DRFs_Nasal_No_Part_Full} shows the DRF contour for VC/CV segments of English words uttered by different male and female speakers of English, to illustrate the difference in extent of oral--nasal coupling. 
The different VC/CV pairs in the figures are as follows: Fig. \ref{figure:DRFs_Nasal_No_Part_Full}(a1) for \emph{/}\textipa{me}\emph{/} (word `\emph{\underline{me}lody'}, gender: male), Fig. \ref{figure:DRFs_Nasal_No_Part_Full}(b1) for \emph{/}\textipa{m2}\emph{/} (word `\emph{\underline{mo}nday'}, gender: male), Fig. for \ref{figure:DRFs_Nasal_No_Part_Full}(c1) \emph{/}\textipa{\ae n}\emph{/} (word `\emph{\underline{an}'}, gender: female), Fig. \ref{figure:DRFs_Nasal_No_Part_Full}(d1) for \emph{/}\textipa{m2}\emph{/} (word `\emph{le\underline{mo}n'}, gender: female), Fig. \ref{figure:DRFs_Nasal_No_Part_Full}(e1) for \emph{/}\textipa{mi}\emph{/}(word `\emph{\underline{me}'}, gender: male) and Fig. \ref{figure:DRFs_Nasal_No_Part_Full}(f1) for \emph{/}\textipa{ni}\emph{/} (word `\emph{mo\underline{ney}'}, gender: male). 
The VC/CV transition boundary can be located around $50$ ms for all the segments as given by respective annotations. 
In Figs. \ref{figure:DRFs_Nasal_No_Part_Full}(a2) and \ref{figure:DRFs_Nasal_No_Part_Full}(b2), the DRF contours corresponding to oral and nasal segments appear bounded within the $B_V$ and $B_N$ range, respectively. 
Figs. \ref{figure:DRFs_Nasal_No_Part_Full}(c2) and \ref{figure:DRFs_Nasal_No_Part_Full}(d2) show the cases where the DRF contours for vowel segments, fluctuate between the $B_V$ and $B_N$ range. 
This reflects the simultaneous dominance of oral and nasal resonances, and hence the vowel is hypothesized as partially nasalized. 

Figs. \ref{figure:DRFs_Nasal_No_Part_Full}(e2) and \ref{figure:DRFs_Nasal_No_Part_Full}(f2) show the DRF contours 
appearing completely in the $B_N$ range for the vowel segment, with a behavior similar to nasal segment. 
The VC/CV transition boundaries occur around $50$ ms  (Figs. \ref{figure:DRFs_Nasal_No_Part_Full}(e1) and \ref{figure:DRFs_Nasal_No_Part_Full}(f1)) , which cannot be observed in the respective DRF contours. 
This behavior reflects the presence of a dominant nasal resonance for the entire duration of vowel segment, which is attributed to a larger extent of coupling of oral and nasal tracts. 
Such a coupling masks oral resonance, and hence the DRF contours shift completely towards $B_N$ range. 
The following Secs. \ref{section:sec4.2} and \ref{section:sec4.4}  present further discussions on this. 

\subsection{Significance of glottal open region for studying degree of nasalization}
\label{section:sec4.2}
Relation between glottal open phase and degree of nasalization, with a higher open quotient leading to a higher degree of nasalization, has been studied in literature \cite{chen1995acoustic}. 
It has also been reported that spectral behavior for a nasalized vowel is similar to addition of the transfer function of oral and nasal cavities, along with a shift and broadening of the formant peaks \cite{stevens2000acoustic}. 
The present section examines these factors, based on the behavior of DRF contour in nasalized vowel segments.  

\begin{figure}
\centering
  \begin{tabular}{@{}c@{}}
    \includegraphics[height = 6cm, width=10cm, trim=0cm 0cm 0cm 0cm]{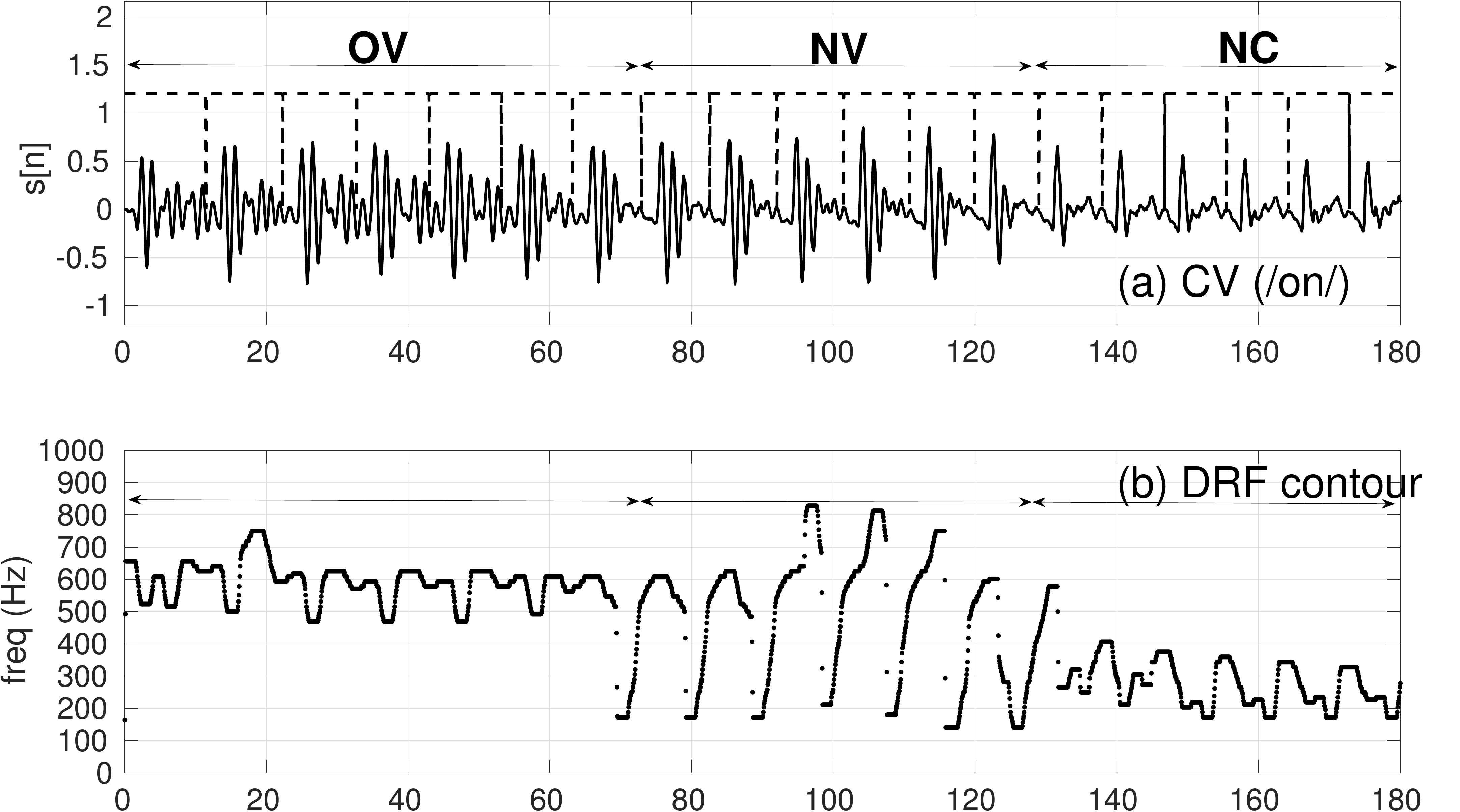}   \\ \\
%     \caption{ABC}
    \includegraphics[height = 10cm, width=12cm, trim=0cm 1cm 0cm 0cm]{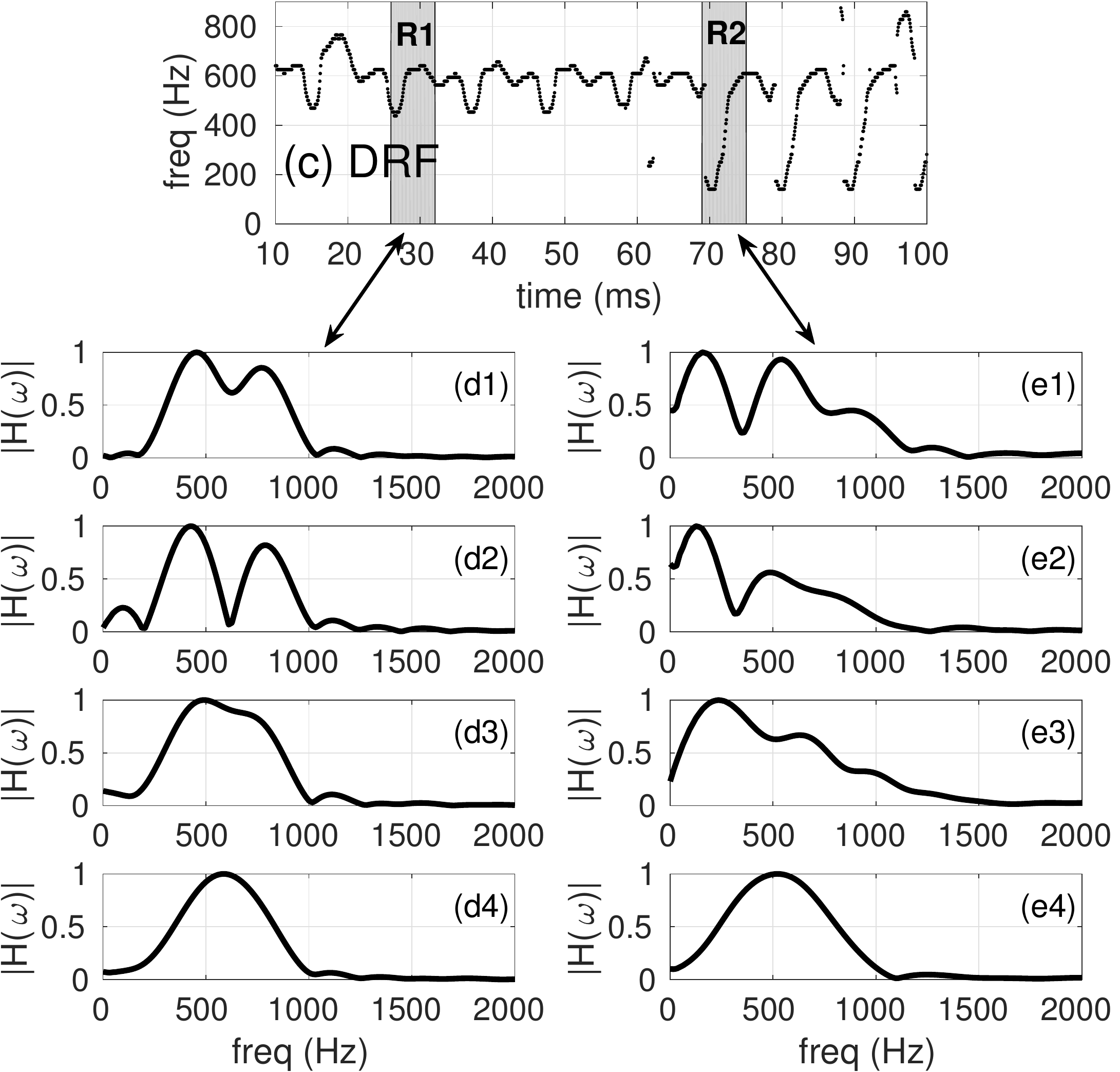}\\ \\
  \end{tabular}
\caption{Illustration of DRF contours and HNGD spectra for the partial opening of the velopharyngeal port during the production of the vowel segment. 
(a) speech signal for VC /on/ (word: \underline{on}ly) with GCI. (b) DRF contour. 
(c) Magnified DRF contour from $10$--$100$ ms with analysis regions R$1$ and R$2$.
(d1)--(d4) HNGD spectra at equidistant windows in region R$1$.
(e1)--(e4) HNGD spectra for equidistant windows in region R$2$
}
\label{figure:DRF1_2and2}
\end{figure}

Fig. \ref{figure:DRF1_2and2} illustrates a case of partial extent of nasalization of a vowel segment. 
Once again, this claim of extent is based on the behavior of DRF contour which fluctuates between $B_V$ and $B_N$ range. 
Fig. \ref{figure:DRF1_2and2}(a) shows the speech signal along with the GCI locations for the VC segment (/o\textipa{n}\emph{/} word: \emph{`\underline{on}ly'} in English language, uttered by English male speaker). 
Fig. \ref{figure:DRF1_2and2}(b) shows the DRF contour obtained from the HNGD spectra with $l=6$ ms. 
The regions of nasal consonant (NC), nasalized vowel (NV) and oral vowel (OV) segments are marked manually across the CV segment based on the DRF contour behavior, with NV exhibiting a partial degree of nasalization in $70$--$130$ ms duration. 
The $B_V$ and $B_N$ ranges can be identified from OV and NC segments in range $500$--$650$ Hz and $200$--$400$ Hz, respectively. 
A close observation suggests that the DRFs in vicinity of GCIs appear in the $B_V$ range, whereas those in the middle of two GCI locations appear transit to the $B_N$ range. 
%%%%%%%%%%%%%%%%%%%%%%%%%%%%%%%%%%%%%%%%%%%%%%%%%%%%%%%%
%%%%%%%%%%%%%%%%%%%%%%%%%%%%%%%%%%%%%%%%%%%%%%%%%%%%%%%%
%%%%%%%%%%%%%%%%%%%%%%%%%%%%%%%%%%%%%%%%%%%%%%%%%%%%%%%%

Fig. \ref{figure:DRF1_2and2}(c) shows a magnified view of the DRF contour for the segment from OV and NV regions, in $10$--$100$ ms duration. 
The figure also marks two regions $R1$ and $R2$ within OV and NV regions, demarcated manually, which essentially span from the onset of the glottal open phase to part of the close phase, as derived by the behavior of DRF contour \cite{rspJASA}. 
These regions are identified to illustrate the evolution of spectral behavior across a glottal cycle for OV segments, and NV segments owing to a partial degree of nasalization. 
The closed region behavior is nearly identical for both OV and NV segments. 
The spectrum within $R1$ and $R2$, sampled at four equidistant locations, is showed in Figs. \ref{figure:DRF1_2and2}(d1)--(d4) and \ref{figure:DRF1_2and2}(e1)--(e4), respectively. 
Figs. \ref{figure:DRF1_2and2}(d1)--(d2) correspond to glottal open region for R$1$. 
A spectral dominance can be noted around $450$ Hz, which further can be seen shifting to higher frequency range ($\sim 600$ Hz) during the glottal closed phases, in Figs. \ref{figure:DRF1_2and2}(d3)--(d4).
During the glottal open phase for OV segments, this close phase resonance shifts to a higher $700$--$750$ Hz as a secondary peak, in Fig. \ref{figure:DRF1_2and2}(d2). 

A shift in location of oral vowel formants during nasalization has already been discussed in the literature \cite{chen1995acoustic}.  
Figs. \ref{figure:DRF1_2and2}(e1)--(e4) show the spectral evolution during $R2$, present in NV segment. 
Dominance of low frequency characteristic nasal resonance at $200$ Hz during glottal open phase (Figs. \ref{figure:DRF1_2and2}(e1)--(e3)), along with the shifted oral resonance ($ \sim 600$--$800$ Hz) impress upon the coupling of oral and nasal tracts.  
The dominant behavior of the low frequency resonance during the glottal open phase transits back to oral formant location during glottal closed phase, in Fig. \ref{figure:DRF1_2and2}(e4).  
Similarity in DRF during glottal closed phase for OV and NV segments can be noted in Figs. \ref{figure:DRF1_2and2}(d4) and \ref{figure:DRF1_2and2}(e4).  
The oral resonances are weaker and hence masked by presence of a stronger nasal zero in the vicinity of first formant, during low SNR glottal open phase. 
The figure illustrates the alternating dominance of oral and nasal resonances for a partial extent of nasalization. 
This verifies the hypothesis of a partial coupling of oral--nasal tracts, as claimed earlier for Figs. \ref{figure:DRFs_Nasal_No_Part_Full}(c2) and \ref{figure:DRFs_Nasal_No_Part_Full}(d2). 

The effect of analysis window duration on identification of glottal open and closed phases in speech has been discussed in the literature \cite{rspJASA}. 
The dominant resonances during the glottal open phase are weaker and hence are masked if the size of the analysis window is longer than few pitch periods. 
It is therefore necessary to maintain the analysis window duration comparable to the pitch duration to study nasalization in vowels. 

\subsection{Full extent of coupling of oral--nasal tracts}
\label{section:sec4.4}
\begin{figure}
\centering
\includegraphics[height = 7cm, width=\linewidth, trim=0cm 0cm 0cm  5cm]{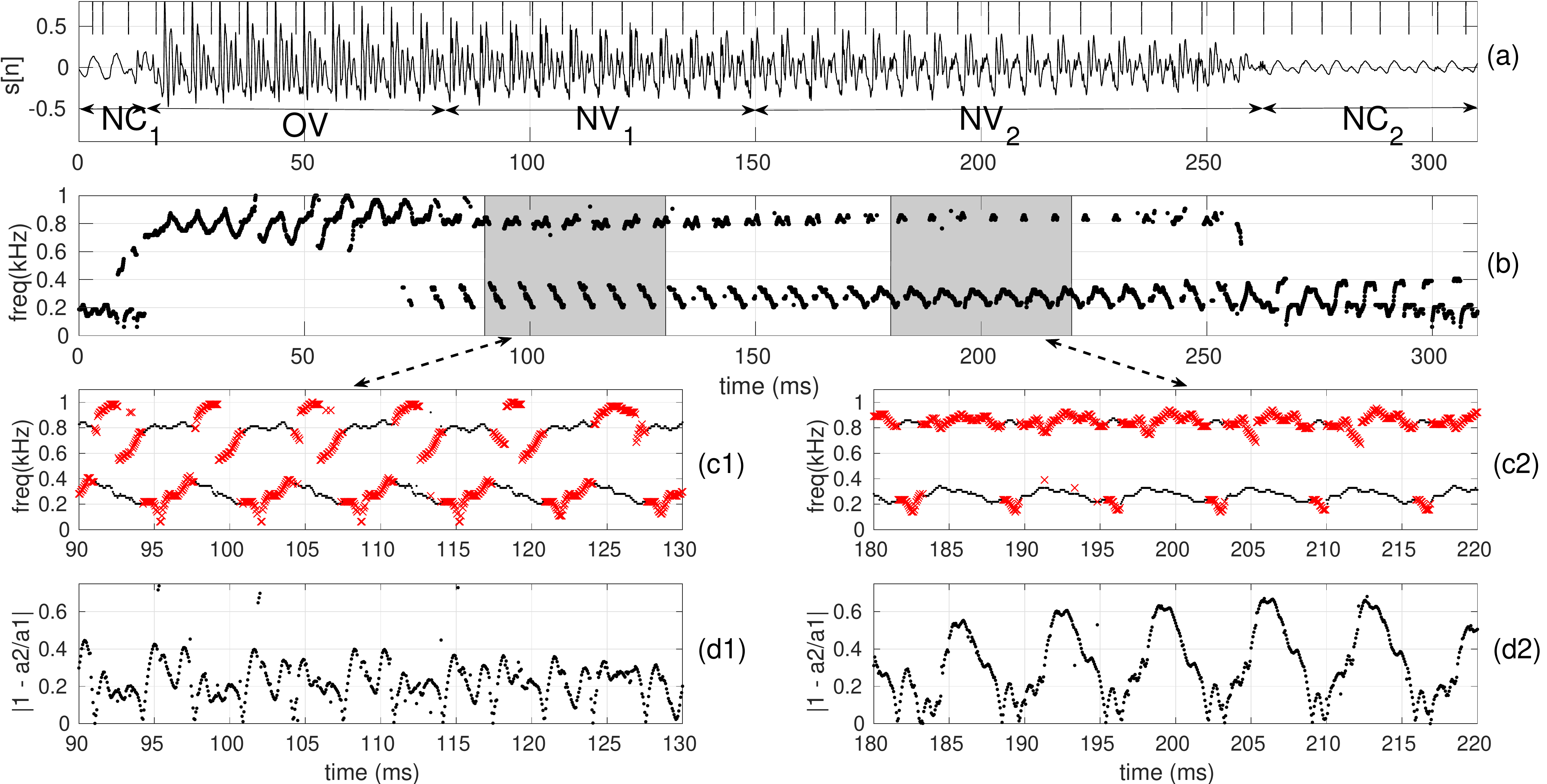} %Figure7
\caption{DRF and DRF$_2$ contours illustrating the variation in degree of coupling of nasal and oral tracts. 
(a) Speech segment with the GCI locations (dotted).
(b) DRF contour. (c1) and (c2) DRF (solid line) along with DRF$_2$ (**).
(d1) and (d2)  $|1 - a2/a1|$.} 
\label{figure:FullNasal}
\end{figure}

It has been discussed in the literature that a higher extent of coupling of oral--nasal tracts has a more drastic effect on the vowel spectra \cite{stevens2000acoustic}. 
A larger opening of the velopharyngeal section results in a stronger nasal resonance which dominates the oral resonance during entire glottal cycle. 
Oral resonance experiences significant decay due to a higher extent of coupling. 
Fig. \ref{figure:FullNasal} illustrates this phenomenon with the help of strength of resonances. 

Fig. \ref{figure:FullNasal}(a) shows a vowel segment \emph{/}\textipa{\ae}\emph{/}, present in context of nasal consonants \emph{/}\textipa{m}\emph{/} and \emph{/}\textipa{n}\emph{/} (word: \emph{`\underline{man}'} in English language by a female English speaker). 
The GCI locations for the segment are shown by dotted vertical lines, along with demarcations for nasal segments (NC$_1$ and NC$_2$), oral vowel (OV) and nasalized segments (NV$_1$ and NV$_2$).  
Fig. \ref{figure:FullNasal}(b) shows the DRF contour for the segment obtained using the ZTW method with $l=6$ ms. 
DRFs in NC$_1$ and NC$_2$, and OV segments appear in $B_N$ and $B_V$ range. 

%%%%%%%%%%%%%%%%%%%%%%%%%%%%%%%%%%%%%%%%%%%%%%%%%%%%%%%%
%%%%%%%%%%%%%%%%%%%%%%%%%%%%%%%%%%%%%%%%%%%%%%%%%%%%%%%%

The NV$_1$ and NV$_2$ segments ($80$--$260$ ms duration) are nasalized. 
Splitting of nasalized vowel segment within NV$_1$ and NV$_2$ is done to illustrate the difference of extent of coupling of oral--nasal tracts. 
Regions within NV$_1$ and NV$_2$ are chosen to illustrate the difference in behavior of DRF contour for different extent of nasalization. 
Figs. \ref{figure:FullNasal}(c1) and \ref{figure:FullNasal}(c2) show the DRF (solid line) and DRF$_2$ (**) contours within NV$_1$ and NV$_2$ segments, respectively. 
The DRF$_2$ are the secondary dominant resonances in the spectrum and are used in this context to illustrate the shift in dominance. 
These are identified as resonances with strength next to DRF in the HNGD spectrum. 
Figs. \ref{figure:FullNasal}(d1) and \ref{figure:FullNasal}(d2) show the relative difference $(\alpha=|1 - (a2/a1)|)$ of amplitudes $a1$ and $a2$ of DRF and DRF$_2$, respectively, within NV$_1$ and NV$_2$ segments, respectively. 

The NV$_1$ segment exhibits a partial extent of nasalization as suggested by the DRF contour. 
The DRF$_2$ contour (Fig. \ref{figure:FullNasal}(c1)) appears appropriately in the gaps in DRF contour, suggesting the simultaneous presence of resonances in $B_V$ and $B_N$ range, across the vowel segment.
DRF contour in NV$_2$ segment shifts completely to $B_N$ range reflecting the increased dominance of nasal resonance over the oral resonance. 
The DRF$_2$ also shifts appropriately to $B_V$ range (Fig. \ref{figure:FullNasal}(c2)). 
Average value of $\alpha$ can also be observed rising with each glottal cycle while transiting from NV$_1$ to NV$_2$ segments (Fig. \ref{figure:FullNasal}(d2)). 
This reflects a stronger nasal resonance compared to the oral resonance in NV$_2$ segments. 
A stronger nasal characteristic resonance within vowel segments signifies a larger extent of velopharyngeal opening.

\section{Observations at CV/VC boundaries for vowel--nasal pairs}
\label{observationCVCboundaries}
The section validates the relation between $\sigma_D$ and $A1-P0$ for different sets of oral and nasalized vowels. 
The section also examines the presence of nasalization owing to co--articulatory load across multiple instances of CV/VC and CVC utterances in English. 

\subsection{Relation in $A$\emph{1}--$P$\emph{0} and $\sigma_D$}
%%%%%%%%%%%%%%%%%%%%%%%%%%%%%%%%%%%%%%%%%%%%%%%%%%%%%%%%%%%%%%%%
\begin{figure}[htb]
\centering
\includegraphics[height = 6.5cm, width=\linewidth]{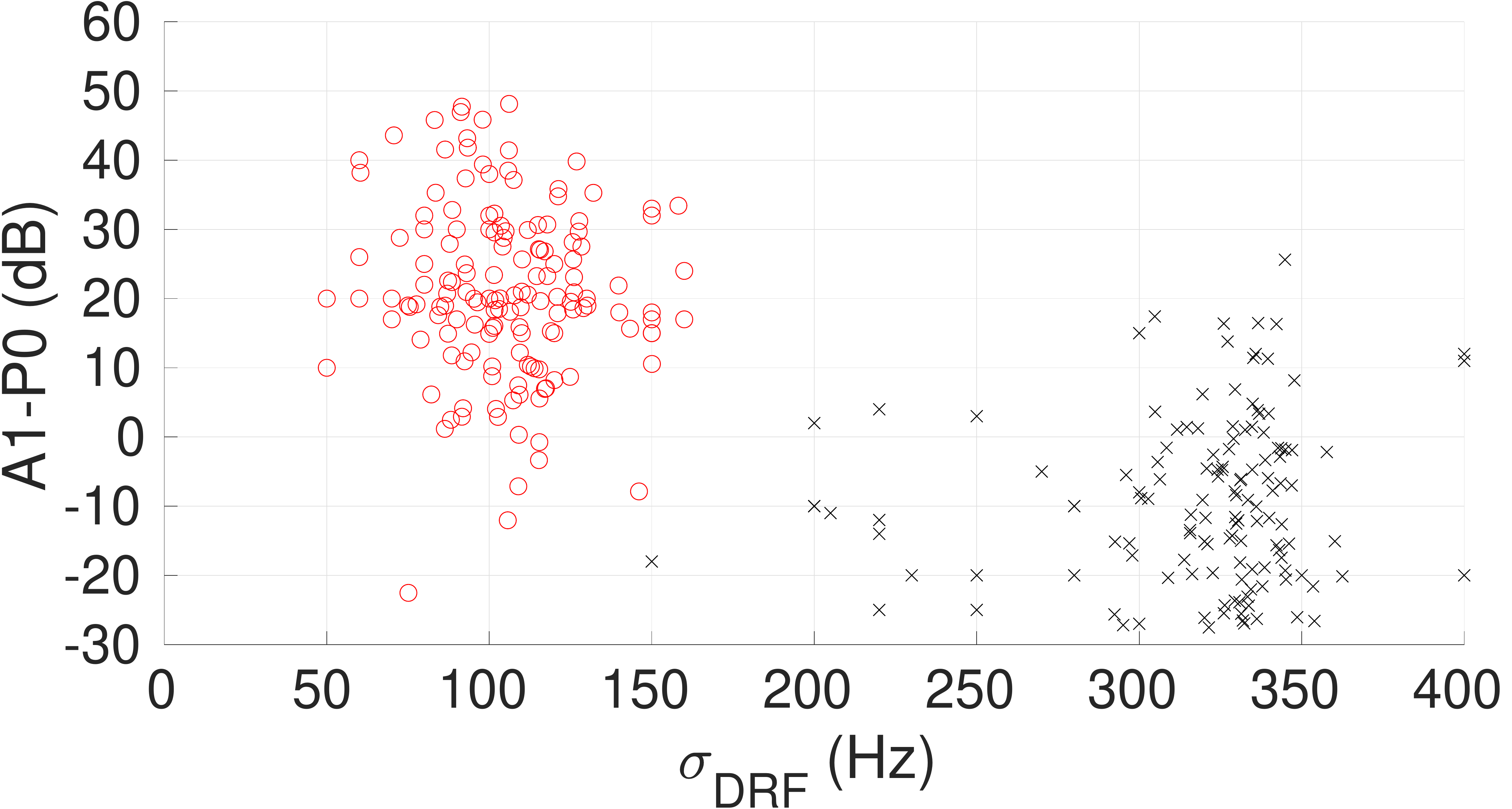}
\caption{$A$\emph{1}--$P$\emph{0} vs $\sigma_{\text{DRF}}$ for OV ({\color{red}$\circ$}) and NV ($\times$) segments. }
\label{figure:fig2poin5}
\end{figure}
%%%%%%%%%%%%%%%%%%%%%%%%%%%%%%%%%%%%%%%%%%%%%%%%%%%%%%%%%%%%%%%% 
Values of $A$\emph{1}--$P$\emph{0}, and their relation with $\sigma_D$ values, is explored for OV and NV segments in TIMIT database \cite{TIMIT}. 
The OV and NV segments are identified using the proposed hypothesis based on the DRF contours. 
Fig. \ref{figure:fig2poin5} gives values obtained for $A$\emph{1}--$P$\emph{0} against $\sigma_D$ for different OV and NV segments obtained across several speakers in TIMIT. 
The OV segments are obtained from vowels appearing in CV/VC clusters with fricative or stop consonants. 
The NV segments are obtained from vowels appearing in context of a nasal consonant with their DRFs fluctuating to $B_N$ range. 
The $A$\emph{1}--$P$\emph{0} values are derived over a $20$ ms frame duration at VC/CV transition region for vowel segments. 
Peaks corresponding to $A$\emph{1} and $P$\emph{0} are marked manually in the normalized log spectrum obtained using DFT. 
$\sigma_D$ values obtained with ZTW using analysis window of $4$ ms, are averaged across a duration of $20$ ms at VC/CV transition boundary. 
It is observed that $A$\emph{1}--$P$\emph{0} values for oral vowel segments ({\color{red}$\circ$}) are higher and positive . 
This signifies a stronger $F_1$ as compared to low frequency resonance. 
The corresponding $\sigma_D$ values are also found to be relatively lower ($\sim 70-130$ Hz). 
The NV segments on the other hand, result in negative $A$\emph{1}--$P$\emph{0} values ($\times$) appearing in distinct cluster. 
Values of $\sigma_D$ are also relatively higher ($\sim 320-360$ Hz) for NV segments. 
These two non--overlapping clusters suggests that proposed analysis aligns with previously suggested spectral correlates to capture nasalization in vowels. 
The proposed method based on DRF contour however, reflects the spectral information at a better resolution, and makes it relatively easier to be characterised. .

\subsection{Study of vowels at VC/CV transition boundaries}
\label{drf_behav}

\begin{figure}[htb]
\centering
\includegraphics[height = 6cm, width=\linewidth]{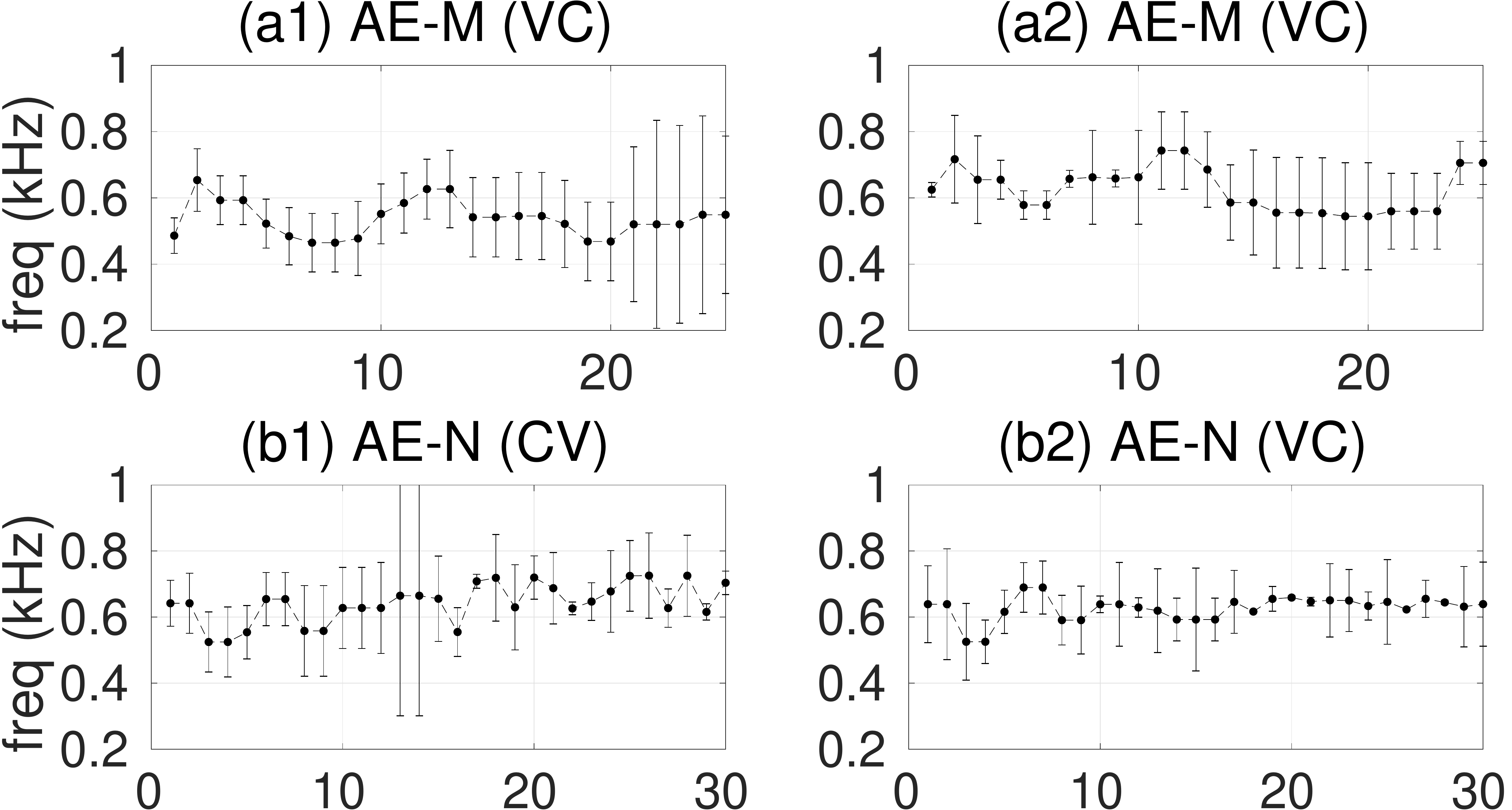}
\vspace{-.2in}
\caption{Errorbars for average $\mu_D$ and $\sigma_D$ values obtained across VC/CV boundaries for vowel /\textipa{\ae}/. }
\label{figure:vowelAE}
\end{figure}

\begin{figure}[!ht]
\centering
\includegraphics[height = 6cm, width=\linewidth]{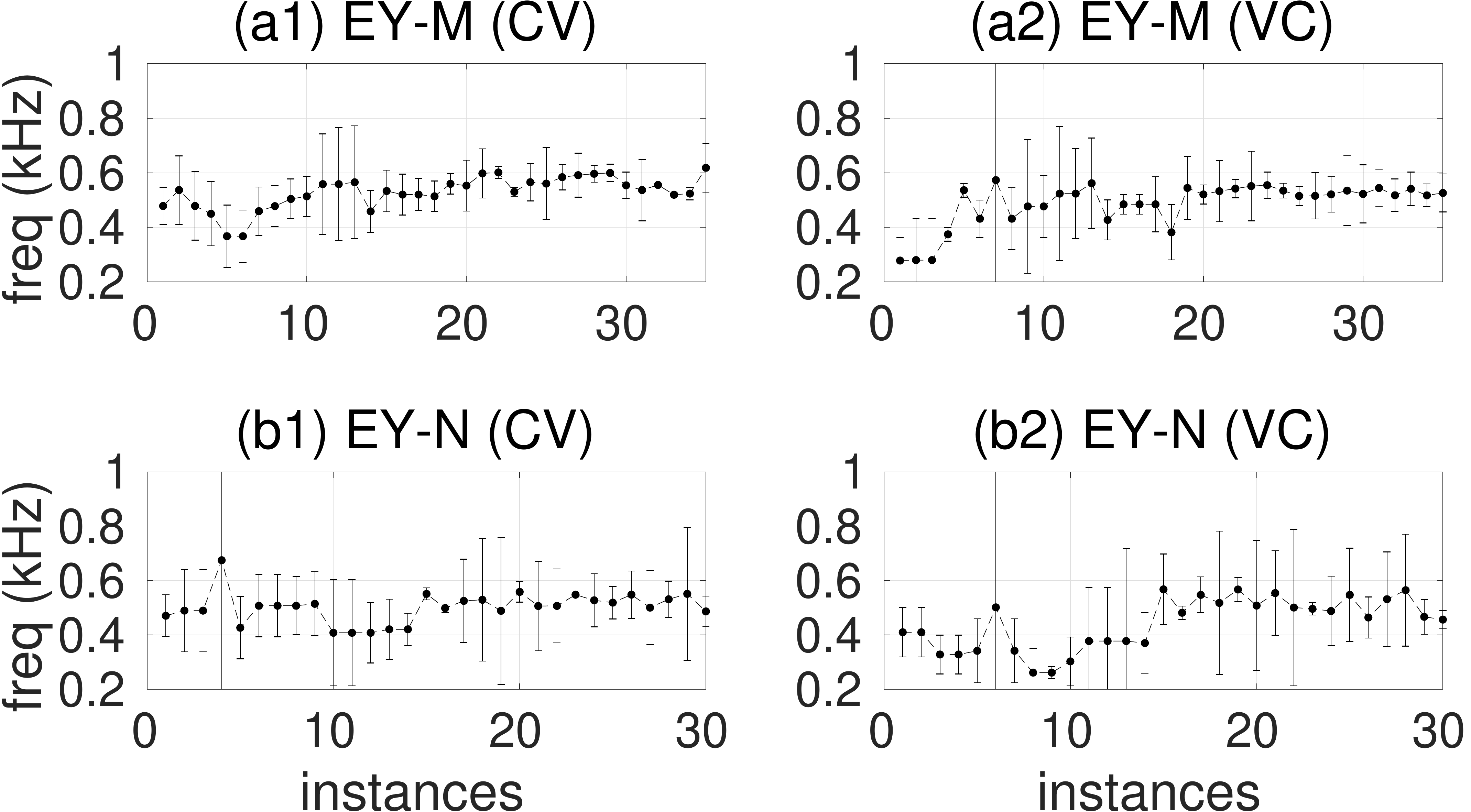}
\caption{Errorbars for average $\mu_D$ and $\sigma_D$ values obtained across VC/CV boundaries for vowel /\textipa{Y}/. }
\label{figure:vowelEY}
\end{figure}

The proposed method is employed to examine the behavior of vowel segment at CV/VC transition boundaries when present in context with nasal consonants. 
The $\mu_D$ and $\sigma_D$ values are derived from DRF contours, obtained using ZTW analysis with $l=5$ms. 
The VC/CV segments are obtained from utterances by different male and female speakers in TIMIT. 
Values of $\mu_D$ and $\sigma_D$, averaged across a duration of $5$ glottal cycles in the vowel segment, at the CV/VC transition boundary are reported. 
Figs. \ref{figure:vowelAE} and \ref{figure:vowelEY} show observations for vowels /\textipa{\ae}/ and /\textipa{Y}/ present in context with nasal consonants /m/ and /n/, in utterances obtained across different speakers.  

%%%%%%%%%%%%%%%%%%%%%%%%%%%%%%%%%%%%%%%%%%%%%%%%%%%%%%%%%%%%%%%%
%%%%%%%%%%%%%%%%%%%%%%%%%%%%%%%%%%%%%%%%%%%%%%%%%%%%%%%%%%%%%%%% 
%%%%%%%%%%%%%%%%%%%%%%%%%%%%%%%%%%%%%%%%%%%%%%%%%%%%%%%%%%%%%%%% 

Few segments exhibit a similarity in behavior for $\mu_D$ values, in /\textipa{\ae}/, where these values are present in the range $B_V$, as seen in Figs. \ref{figure:vowelAE}(a1)--\ref{figure:vowelAE}(b2). 
Some of these values for /\textipa{Y}/, especially for VC transition, appear in range $B_N$ which shows the affinity of this vowel to get nasalized, as seen in Figs. \ref{figure:vowelEY}(a2)--\ref{figure:vowelEY}(b2). 
Few instances for /\textipa{\ae}/ at VC boundary with /m/ (Fig. \ref{figure:vowelAE}(a1)) show increased value of $\sigma_D$, which also illustrates the overlap of $B_V$ and $B_N$ and hence presence of nasalization. 
The figures illustrate a dynamic nature of nasalization due to contextual load across similar CV/VC clusters, for different speakers.

\section{SUMMARY}
\label{summary}
This paper presents a new way to study the phenomena of nasalization of vowels based on dominant behavior of the instantaneous spectral characteristics. 
The problems of identification of duration and extent of nasalization in vowels are addressed in this paper. 
The DRFs obtained from HNGD spectrum fluctuate within characteristic $B_V$ and $B_N$ range for oral vowels and nasal consonants, respectively. 
Changes in the spectral structure of vowels due to coupling of oral--nasal tracts are captured based on the average location and spread of DRFs. 
This behavior is characterized across each glottal cycle using the $\mu_D$ and $\sigma_D$ parameters. 
A partial extent of nasalization examined across glottal open and closed phase, and is illustrated through fluctuations in DRFs within $B_V$ and $B_N$ range. 
The paper also explains the shift in dominance within oral and nasal resonances across glottal phases, for a partial nasalization. 
A higher extent of coupling is illustrated by a stronger nasal resonance dominating across the entire glottal cycle. 
Examination of secondary dominant resonance proves a continuum in DRF contours in $B_V$ and $B_N$ range during nasalization. 
Comparison of strengths of DRF and DRF$_2$ shows the transition of dominance within oral and nasal resonances.   

The study is carried out over nasal--vowel pairs occurring words in the English language, spoken by different male and female English speakers. 
The behavior of $\sigma_D$ for nasalized vowel segments is also validated using spectral correlate $A1-P0$ derived across VC/CV boundaries. 
The DRF$_2$ contour helps to illustrate the simultaneous presence of oral and nasal resonances in the spectrum for nasalized vowels, and the fluctuation in the dominant behavior between the two. 
The study can easily be implemented across different languages and speakers, and can help in studying speech pathology.

%%%%%%%%%%%%%%%%%%%%%%%%%%%%%%%%%%%%%%%%%%%%%%%%%%%%%%%%%%%%%%%%%%%%%%%%%%%%%%%%%%%%%%%%%%%%%%
%%%%%%%%%%%%%%%%%%%%%%%%%%%%%%%%%%%%%%%%%%%%%%%%%%%%%%%%%%%%%%%%%%%%%%%%%%%%%%%%%%%%%%%%%%%%%%
% \section*{\refname}
\footnotesize
\bibliographystyle{ieeetr}
\bibliography{citation-nasalization}
\end{document}